\documentclass[aps,pra,reprint,twocolumn,amsmath,amssymb,citeautoscript,longbibliography]{revtex4-1}
\usepackage{graphicx}
\usepackage{dcolumn}
\usepackage{bm}
\usepackage{latexsym,epsfig}
\usepackage{graphicx}
\usepackage{verbatim}
\usepackage{comment}
\usepackage{amsmath}
\usepackage{amssymb}
\usepackage{physics}
\usepackage{stmaryrd}
\usepackage{color}
\usepackage{epstopdf}
\usepackage{grffile}
\usepackage{lipsum}
\usepackage{enumitem}
\usepackage{ulem}
\usepackage{hyperref}
\usepackage{cleveref}
\usepackage[dvipsnames]{xcolor}
\DeclareGraphicsExtensions{.eps}

\newcommand{\beq}{\begin{equation}}
\newcommand{\eeq}{\end{equation}}
\newcommand{\bea}{\begin{eqnarray}}
\newcommand{\eea}{\end{eqnarray}}
\newcommand{\ben}{\begin{eqnarray*}}
\newcommand{\een}{\end{eqnarray*}}
\newcommand{\bfig}{\begin{figure}}
\newcommand{\efig}{\end{figure}}

\usepackage{hyperref}
\hypersetup{
    colorlinks=true,      
    urlcolor=blue,
    citecolor=blue,
    linkcolor=blue
}

\usepackage{booktabs}
\usepackage{array}
\usepackage{lipsum} 
\usepackage{array} 

\begin{document}
\title{Flux-driven charge and spin transport in a dimerized Hubbard ring with Fibonacci modulation} 

\author{Souvik Roy$^{1,2*}$, Soumya Ranjan Padhi$^{1,2}$ and Tapan Mishra$^{1,2}$}

\affiliation{$^1$ School of Physical Sciences, National Institute of Science Education and Research, Jatni 752050, India}

\affiliation{$^2$ Homi Bhabha National Institute, Training School Complex, Anushaktinagar, Mumbai 400094, India}

\email{souvikroy138@gmail.com}

\date{\today}

\begin{abstract}
We study quantum transport in a one-dimensional Hubbard ring with dimerized nearest-neighbor hoppings and a Fibonacci-modulated onsite potential. 
For non-interacting case our analysis reveals that at half-filling, the charge current along with the Drude weight decreases with increasing onsite potential when inter-cell hopping dominates over the intra-cell hopping, while for dominating intra-cell hopping it shows non-monotonic behavior with sharp peak at certain critical modulation strength, indicating enhanced transport. Moving away from half-filling gives rise to re-entrant features in both quantities at fillings associated with Fibonacci numbers.
On the other hand, in spin-imbalanced systems, both spin and charge current shows multiple peaks and re-entrant behavior, tunable via hopping dimerization and filling. Including the on-site Hubbard interaction preserves the re-entrant behavior in current and moreover favors finite transport which is absent in the non-interacting ring. These results reveal rich interplay among Fibonacci modulated potential, electron fillings, hopping dimerization and interaction.

\end{abstract}

\maketitle

\section{Introduction}
\label{sec:intro}

Anderson’s seminal discovery of localization in disordered systems~\cite{r1,r2,r6,r7,kramer1993,Wiersma1997,Schwartz2007} has enhanced our understanding of the transport phenomenon in modern condensed matter physics. This reveals the fact that an infinitesimal disorder in a one-dimensional (1D) lattices leads to the exponential localization of all single-particle electronic states, thereby suppressing quantum transport. While this phenomenon holds for a lattice with random disorder, deterministic quasiperiodic modulations introduce a contrasting paradigm, enabling fine-tuned control over localization transitions and transport. Among such models, the Aubry-Andr\'e-Harper (AAH) lattice~\cite{r3,rn1,r4,r5} stands as a paradigmatic example, exhibiting a sharp localization-delocalization transition at a critical modulation amplitude, thereby offering a fertile ground to explore metal-insulator transitions without relying on statistical randomness. On the other hand, systems of mesoscopic 1D rings threaded by magnetic flux constitute  well known set ups to understand quantum coherence and transport behavior in the forms of charge currents. Combining the quasiperiodic potentials with Aharonov-Bohm (AB) flux in such geometries opens up a versatile platform to investigate how the interplay between correlated disorder and geometric phase effects influence current-carrying states~\cite{r8,r9,r10,r11,r12,r13,r14}. Notably, within the B\"uttiker-Imry-Landauer framework, the persistence of current even after the withdrawal of the external flux underscores the fundamentally dissipationless and coherent nature of such transport, making the flux-driven quasiperiodic rings a promising architecture to unearth unconventional transport regimes and engineer tunable quantum functionality.

Parallel advances have focused on quasiperiodic systems, particularly those based on the Fibonacci sequence, where the inherent aperiodicity induces rich modifications to electronic and transport characteristics. These systems serve as an ideal platform to explore regimes that interpolate between crystalline order and complete disorder. Of particular interest is the one dimensional AAH model with hopping dimerization in the nearest neighbour (NN) bonds of the lattice
The interplay between quasiperiodic modulation and dimerized hopping introduces intricate physical behavior, including reentrant transitions and nontrivial localization dynamics, thereby enriching the phenomenology of one-dimensional quasiperiodic systems.

In the framework of the diagonal AAH model, earlier studies have shown that the charge current generally diminishes with increasing quasiperiodic potential strength, leading to a transition at a critical point~\cite{r22,r23,r24,r25,r26,r27,r28}. However, under particular conditions, especially when correlations exist between site energies and hopping dimerization, the current may exhibit non-monotonous behavior, with enhancements occurring despite increasing potential strength. This behaviour becomes especially pronounced at selected fillings, revealing unconventional pathways to modulate transport in quasiperiodic systems~\cite{r29,r30,r31,r32,r33,r34,r35,r36,r37,r38,r39,r40}. Such findings have spurred growing interest in this context by uncovering emergent and tunable transport phenomena beyond standard localization scenarios.

In this work, we explore quantum transport of a system of spinful fermions in a ring lattice threaded by an AB flux by allowing on site modulation according to the Fibonacci sequence. By imposing dimerized NN hopping amplitudes and inter-component interaction, we show that at zero-temperature, the amplitude of charge current are highly sensitive to hopping dimerization, onsite potential modulation, Hubbard interaction strength, and electron filling. 
In the non-interacting case, a pronounced peak in the current appears at a specific onsite potential strength for $t_1 > t_2$ at half-filling, while at certain special fillings, the current initially decreases and then increases, exhibiting a re-entrant behavior with increase in the onsite potential modulation. In the interacting case, as $U$ increases, the peak position of the current systematically shifts with the onsite potential strength. However, for a fixed onsite potential modulation, the current increases from an initially vanishing value, reaches a maximum and decreases again as a function of Hubbard interaction strength.  
Introducing spin-imbalance in the system, multiple such peaks in the current appear, and their positions can be effectively tuned by adjusting the hopping dimerization, the onsite potential modulation and the Hubbard interaction, demonstrating a high degree of controllability in the transport characteristics.
These findings reveal a complex yet tunable landscape governed by aperiodicity, strong correlations, and magnetic flux, opening promising avenues for engineering quantum devices with controllable transport properties in deterministic aperiodic systems~\cite{r58,r59,r60,r61,r62}.

The structure of this paper is organized as follows: Sec.~I introduces the central research problem and objectives. Sec.~II and III establishes the model and theoretical framework using a tight-binding formalism, detailing the system Hamiltonian and the expressions for calculating charge current and Drude weight. In Sec.~IV, we present results for the non-interacting and interacting regime, highlighting how Fibonacci potential modulation, electron fillings, and hopping dimerization collectively influence the transport properties. Sec.~V summarizes the main findings and discusses possible directions for future research.

\section{Model}
\label{sec:model}

\label{Sec1}

\begin{figure}[t]
\centering
\includegraphics[width=1.0\columnwidth]{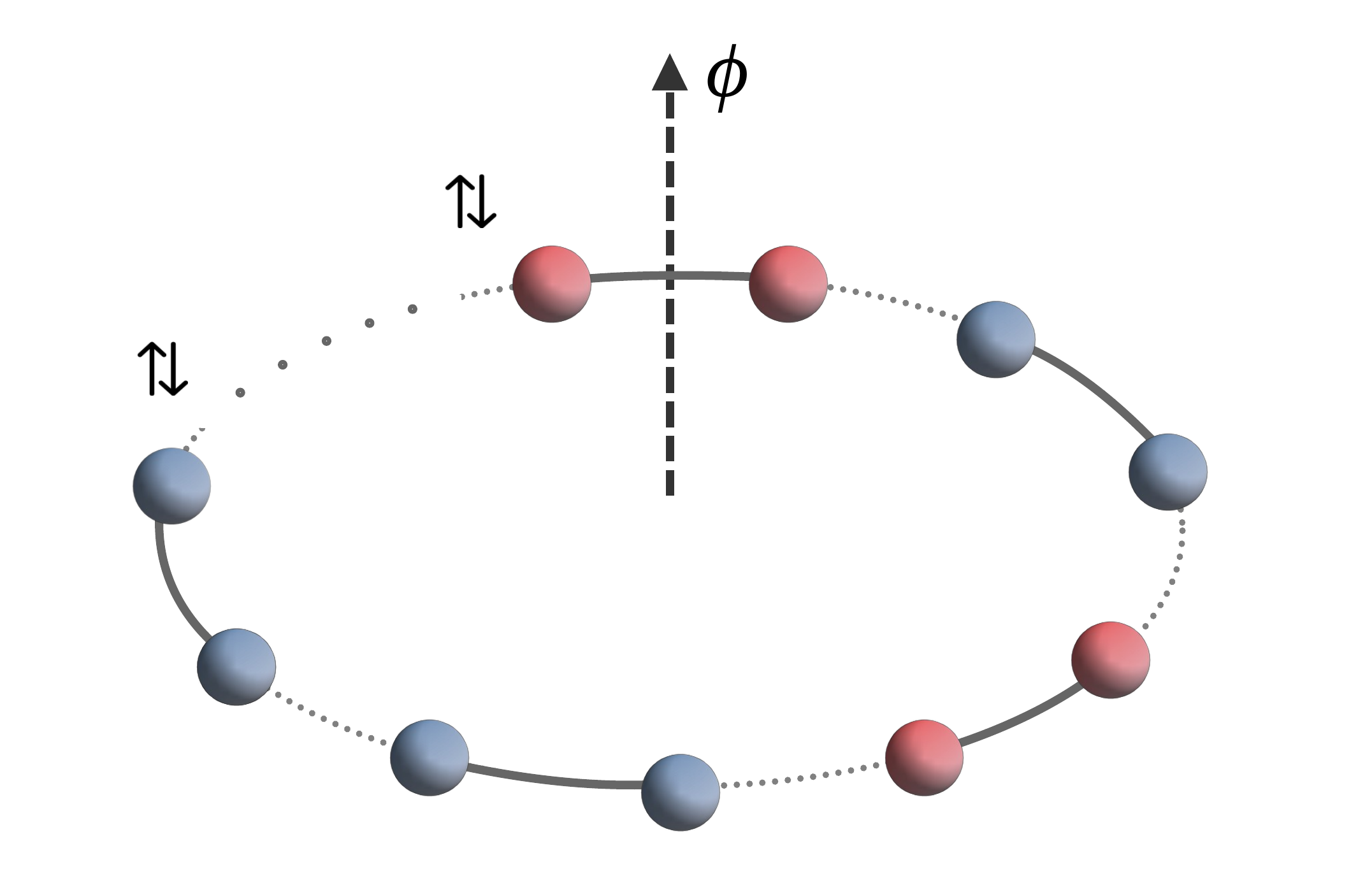}
\caption{The schematic of a one-dimensional ring threaded by an AB flux $(\phi)$, with on-site potentials arranged in a Fibonacci sequence. The system has two types of sites, A (red) and B (blue). Hopping between sites is shown as solid lines for $t_1$ and dashed lines for $t_2$.}
\label{fig:fig1}
\end{figure}

The Hamiltonian that describes a system of spinful fermions on a ring lattice with NN hopping dimerization threaded by flux and onsite potential modulated according to the Fibonacci sequence  is given by 
\begin{align}
    \mathcal{H} =& \sum_{\sigma=\uparrow,\downarrow} \Bigl( 
        \sum_{j\in \text{even}} \big( t_1 e^{i\Phi} c^\dagger_{j,\sigma} c_{j+1,\sigma} + H.c. \big) \label{equation1}\\ \nonumber
        &+ \sum_{j\in \text{odd}} \big( t_2 e^{i\Phi} c^\dagger_{j,\sigma} c_{j+1,\sigma} + H.c. \big) \\ \nonumber
        &+ \sum_{j\in \text{even}} \lambda_{j,\sigma} n_{j,\sigma} 
        + \sum_{j\in \text{odd}} \lambda_{j,\sigma} n_{j,\sigma} 
    \Bigr) + \sum_{j} U n_{\uparrow,j} n_{\downarrow,j}.
\end{align}
where the hopping amplitudes alternate between $t_1$ on even bonds and $t_2$ on odd bonds for both the spin components $\sigma=\uparrow,\downarrow$, representing hopping dimerization.  $c_{j,\sigma}^\dagger$ and $c_{j,\sigma}$ respectively, are the creation and annihilation operators for electron with spin $\sigma = \uparrow, \downarrow$ at site $j$, while $n_{j,\sigma} = c_{j,\sigma}^\dagger c_{j,\sigma}$ denotes the corresponding number operator. The term $\Phi$ in the exponent $e^{i\Phi}$ represents the Peierls phase arising from an AB flux $\phi$ penetrating the ring, where $\Phi = 2\pi\phi/2L\phi_0$, $\phi_0$ denotes the flux quantum, and $\phi$ is expressed in units of $\phi_0$.
The onsite potential $\lambda_{j,\sigma}$ follows a deterministic modulation based on the Fibonacci substitution rule, constructed recursively with $F_0 = B$, $F_1 = A$, $F_2 = AB$, $F_3 = ABA$, and so on. Here, $F_n$ is the Fibonacci number, $A = \lambda_A$ and $B = \lambda_B$ denote the two potential values. This potential sequence is applied uniformly to both even and odd sublattice sites, encoding aperiodic potential pattern throughout the system. The final term represents the onsite Hubbard interaction $U$, accounting for electron-electron repulsion between opposite spins on the same site. We set $t_2=1.0$ as our energy unit.

Within this framework, we aim to elucidate the interplay of AB flux, Fibonacci sequence, dimerized hopping, and Hubbard interaction that shape the quantum transport characteristics of correlated electrons in a one-dimensional ring. To this end we investigate the transport properties by analysing the charge current and Drude weight as key diagnostics. While the charge current reflects the sensitivity of the ground state energy to the AB flux, 
the Drude weight quantifies the charge stiffness and distinguishes between conducting and insulating phases. Note if either $\lambda_A$ or $\lambda_B$ is varied while keeping the other fixed, the underlying physics remains unchanged. Moreover, fixing one of them at a specific value and varying the other primarily leads to a shift or scaling of the peak positions. Therefore, for simplicity, we set one of them to zero and vary the other throughout our analysis.

\section{Methodology}
To analyze the spectral and transport properties of the model shown in Eq.~\ref{equation1}, it is essential to compute the energy eigenvalues corresponding to both spin-up and spin-down electrons. This can be achieved by exact diagonalization (ED) which yields a fully accurate solution by retaining the complete many-body Hilbert space. However, its applicability is severely constrained by the exponential growth of the Hilbert space with system size due to increased computational complexities. Consequently, ED becomes impractical for studying large or even moderately sized systems, particularly when spin and spatial degrees of freedom are both involved. 
While, sophisticated numerical methods like DMRG and quantum Monte-Carlo methods are known to provide quantitatively accurate results for these systems, an efficient alternative is provided by the Hartree-Fock (HF) mean-field approximation, which offers a computationally tractable route to explore the qualitative behavior of interacting systems across larger system sizes. In the following we briefly discuss this method for completeness. 

\label{sec:theory}


\subsection{Mean-field decoupling of the interacting tight-binding Hamiltonian}

Within the MF framework, the interaction term is approximated by replacing the product of number operators with their mean-field decoupled form~\cite{MAITI20102212}. Specifically, the interaction term $U n_{j,\uparrow} n_{j,\downarrow}$ is treated using the HF decoupling scheme, such that
\begin{equation}
U n_{j,\uparrow} n_{j,\downarrow} \approx U \left( \langle n_{j,\uparrow} \rangle n_{j,\downarrow} + \langle n_{j,\downarrow} \rangle n_{j,\uparrow} - \langle n_{j,\uparrow} \rangle \langle n_{j,\downarrow} \rangle \right),
\end{equation}
thereby transforming the two-body interaction into an effective single-particle potential dependent on the local spin densities. As a result, the interacting Hamiltonian is decoupled into two separate spin-dependent single-particle Hamiltonians, which can be diagonalized independently. This mean-field treatment preserves the spin-resolved physics while enabling the efficient computation of energy spectra and the physical quantities of interest.

Within the Hartree-Fock MF framework, the on-site interaction term introduces spin-dependent modifications to the original site energies, resulting in an effective potential experienced by each spin species. Specifically, the effective site energies for up-spin electrons are given by
\begin{equation}
V_{j^{\text{even/odd}},\uparrow}^{\text{eff}} = V_{j,\uparrow} + U\langle n_{j,\downarrow} \rangle,
\end{equation}
where $V_{j,\uparrow}$ denotes the original site energy for the up-spin electron at site $j$, and $\langle n_{j,\downarrow} \rangle$ is the expectation value of the local down-spin electron density. The term $U\langle n_{j,\downarrow} \rangle$ captures the mean-field contribution from the on-site Coulomb repulsion due to the presence of a down-spin electron. This implies that the up-spin electron feels an enhanced potential at each site depending on the local occupation by down-spin electrons.

Analogously, the effective site energies for down-spin electrons can be expressed as
\begin{equation}
V_{j^{\text{even/odd}},\downarrow}^{\text{eff}} = V_{j,\downarrow} + U\langle n_{j,\uparrow} \rangle,
\end{equation}
illustrating that the local environment of a down-spin electron is influenced by the spin-up density through the same interaction mechanism. These modified site energies incorporate the essential effects of electron-electron interactions and self-consistently encode the coupling between spin degrees of freedom via the local densities.

Utilizing these effective potentials, the original interacting Hamiltonian can be recast in a decoupled form suitable for numerical diagonalization,
\begin{equation}
H = H_{\uparrow} + H_{\downarrow} - \sum_j U \langle n_{j,\uparrow} \rangle \langle n_{j,\downarrow} \rangle,
\label{Eq 3}
\end{equation}
where $H_{\uparrow}$ and $H_{\downarrow}$ denote the effective single-particle Hamiltonians for the up- and down-spin electrons, respectively. The final term subtracts the double-counting of interaction energy already included in the mean-field decomposition. This transformation reduces the full many-body problem into two spin-resolved non-interacting problems, where each spin species evolves independently in the presence of a self-consistently determined background potential. Physically, this reflects the indirect coupling between up- and down-spin electrons mediated through the local density fields, preserving key interaction effects while offering significant computational tractability.

The effective up and down spin Hamiltonians become,
\begin{align}
H_{\uparrow} &= \sum_j V_{j,\uparrow}^{\text{eff}} c_{j,\uparrow}^\dagger c_{j,\uparrow} + \sum_j \left( t_{j}^{\uparrow} e^{i\Phi} c_{j,\uparrow}^\dagger c_{j+1,\uparrow} + \text{H.c.} \right), \label{eq:up_ham}
\\
H_{\downarrow} &= \sum_j V_{j,\downarrow}^{\text{eff}} c_{j,\downarrow}^\dagger c_{j,\downarrow} + \sum_j \left( t_{j}^{\downarrow} e^{i\Phi} c_{j,\downarrow}^\dagger c_{j+1,\downarrow} + \text{H.c.} \right).
\label{eq:down_ham}
\end{align}
Here, $t_j^{\uparrow}$ and $t_j^{\downarrow}$ represent the spin-dependent hopping amplitudes, which alternate based on the parity of the site index; specifically, $t_1$ is assigned to even sites and $t_2$ to odd sites.
These expressions represent spin-dependent interaction effects within the mean-field approximation.


To achieve a self-consistent mean-field solution, the procedure begins by initializing the local spin-resolved electron densities with reasonable trial values, typically denoted as $\langle n_{j,\uparrow} \rangle$ and $\langle n_{j,\downarrow} \rangle$. These initial estimates are then used to construct the effective single-particle Hamiltonians $H_{\uparrow}$ and $H_{\downarrow}$ for the up- and down-spin electrons, incorporating the corresponding spin-dependent potentials derived from the mean-field decoupling of the interaction term.
Subsequently, these decoupled Hamiltonians are diagonalized to obtain updated eigenstates and eigenvalues, from which the new local densities are computed. The effective potentials are then recalculated using these updated densities, and the process is repeated iteratively. After each iteration, the changes in the local densities are monitored, and the loop continues until the variation in $\langle n_{j,\uparrow} \rangle$ and $\langle n_{j,\downarrow} \rangle$ between successive steps falls below a predefined convergence threshold, typically set to a small number (e.g., $10^{-5}$ or lower) to ensure numerical stability and physical accuracy.  We choose system sizes according to Fibonacci sequence to capture the essential physics. Unless specified, we consider a 1D dimerized ring of $F_{10}=144$  unit cell (i.e., $288$ sites).

This iterative self-consistent cycle ensures that the final spin densities are in equilibrium with the effective fields generated by the interactions themselves. The converged densities thus represent the stable mean-field configuration of the interacting many-body system, capturing the essential influence of electron-electron correlations within the limitations of the Hartree-Fock approximation. 

\subsection{Ground State Energy and Current}
At zero temperature, the ground-state energy is governed by the sum of all occupied single-particle eigenvalues up to the Fermi level, leading to  
\begin{equation}
E_0 = \sum_{j=1}^{n_{\uparrow}} E_{j,\uparrow} + \sum_{j=1}^{n_{\downarrow}} E_{j,\downarrow} 
- \sum_j U \langle n_{j,\uparrow} \rangle \langle n_{j,\downarrow} \rangle,
\end{equation}
where $E_{j,\uparrow}$ ($E_{j,\downarrow}$) denotes the eigenenergy of the $j^{\text{th}}$ up (down) spin state, and $n_{\uparrow}$ ($n_{\downarrow}$) is the number of up (down) spin electrons determined by the filling.  
The persistent or charge current is obtained from the flux derivative of the ground-state energy,  
\begin{equation}
I_c = -c \, \frac{\partial E_0}{\partial \phi},
\end{equation}
where $c$ is a constant.
In contrast, the persistent spin current originates from the spin-resolved contributions, and can be evaluated directly as  
\begin{equation}
I_s = I_{\uparrow} - I_{\downarrow},
\end{equation}
with $I_{\uparrow}$ and $I_{\downarrow}$ obtained from the flux response of $E_{j,\uparrow}$ and $E_{j,\downarrow}$, respectively.

In order to analyse the eigenstate dependent transport, we compute the state currents along the lattice. The current operator ($\hat{I}$) is defined as
\[
\hat{I} = \frac{e\hat{v}}{L},
\]
where, $\hat{v}$ denotes the velocity operator and $L$ is the system length. 
In quantum theory, $\hat{v}$ can be expressed through the commutator of the position operator with the Hamiltonian,
\[
\hat{v} = \dot{\hat{x}} = \frac{1}{i\hbar}[\hat{x}, H].
\]
Within this framework, the position operator reads as
\[
\hat{x} = \sum_{j,\sigma} j\, c_{j,\sigma}^\dagger c_{j,\sigma},
\]
with $c_{j,\sigma}^\dagger$ ($c_{j,\sigma}$) creating (annihilating) an electron of spin $\sigma$ at site $j$. 
Substituting this into the commutator with the model Hamiltonian yields
\begin{align}
    \hat{I_{\sigma}} &= \frac{2\pi e}{i\hbar\,2L} 
    \bigg[\sum_{\substack{j \in \text{even}\\}} \left(t_1 e^{i\Phi} c_{j+1,\sigma}^\dagger c_{j,\sigma} - t_1 e^{-i\Phi} c_{j,\sigma}^\dagger c_{j+1,\sigma} \right) \nonumber \\
    &\quad + \sum_{\substack{j \in \text{odd}\\}} \left(t_2 e^{i\Phi} c_{j+1,\sigma}^\dagger c_{j,\sigma} - t_2 e^{-i\Phi} c_{j,\sigma}^\dagger c_{j+1,\sigma} \right) \bigg],
\end{align}
where $\phi$ represents the magnetic flux and $t_{1}$, $t_{2}$ are alternating hopping amplitudes.
The current corresponding to an eigenstate $\lvert \psi_k \rangle$ is obtained by evaluating the expectation value of the current operator,
\[
\langle \psi_k \lvert \hat{I}_{\sigma} \rvert \psi_k \rangle.
\]
The eigenstate $\lvert \psi_k \rangle$ admits the following expansion in the Wannier basis,
\begin{equation}
\lvert \psi_k \rangle = \sum_{j,\sigma} a_{j,\sigma}^{(k)} \lvert j,\sigma \rangle,
\label{eq:eigenstate_expansion_compact}
\end{equation}
where $a_{j,\sigma}^{(k)}$ are the expansion coefficients corresponding to site $j$, spin $\sigma \in \{\uparrow,\downarrow\}$, and eigenstate index $k$. The current for this eigenstate can then be written as
\begin{align}
I_{\sigma}^{(k)} &= \frac{ie}{2L\hbar} \sum_{\sigma=\uparrow,\downarrow} 
\Bigg[ \sum_{j\,\text{odd}} t_1 
\Big( e^{-i\Phi}\, a_{j,\sigma}^{(k)\ast}\, a_{j+1,\sigma}^{(k)} - \text{h.c.} \Big) \nonumber \\
&\qquad\qquad\qquad + \sum_{j\,\text{even}} t_2 
\Big( e^{-i\Phi}\, a_{j,\sigma}^{(k)\ast}\, a_{j+1,\sigma}^{(k)} - \text{h.c.} \Big) \Bigg].
\label{eq:eigenstate_current_dimerized_compact}
\end{align}

\begin{figure*}[htp]
\centering
\includegraphics[width=2.0\columnwidth]{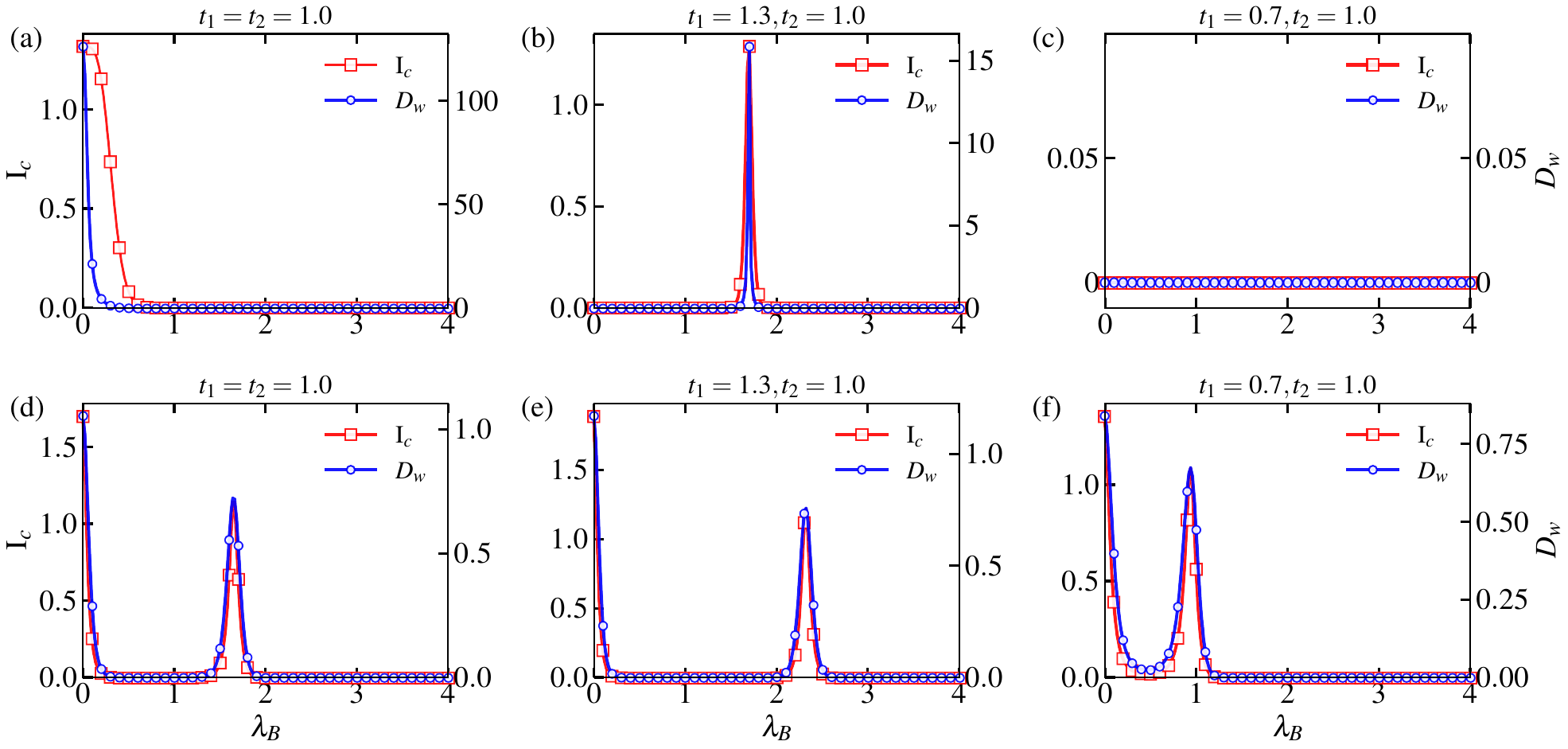}
\caption{The charge current $I_c$ (red squares) and Drude weight $D_w$ (blue circles) are plotted as functions of the onsite potential $\lambda_B$ for different hopping dimerizations. 
Upper panels (a-c) correspond to the half-filling case $(F_n=144)$ 
whereas the lower panel (d-f) depict results for special filling case $(2F_{n-1} + F_{n-4}=199)$ with system size $2F_n=288$. 
The hopping amplitudes are set as $t_1 = 1.0$, $1.3$, and $0.7$, in left, middle and right columns, respectively, with $t_2$ fixed at 1.0 with $\lambda_A=0$.}
\label{fig:fig2}
\end{figure*}
Apart from current, we also calculate the Drude weight ($D_w$) characterizes the charge stiffness and reflects the ability of the system to support dissipationless current flow.  At zero temperature, the Drude weight at zero flux $\phi=0$ characterizes the intrinsic conductivity of the system in the absence of an external magnetic field and is defined as,
\begin{equation}
    D_w = \frac{2L}{4\pi^2} \left. \frac{\partial^2 E_0}{\partial \phi^2} \right|_{\phi=0},
\end{equation}
where $E_0$ is the ground-state energy.

\section{Results}
\label{sec:results}
In the following, we present our results for the non-interacting and interacting regimes in separate subsections for clarity. The non-interacting case is analyzed first, focusing on the effects of potential strength, hopping dimerization, and electronic filling on transport properties. The interacting case incorporates electron-electron correlations through the on-site Hubbard interaction \(U\), studied with the same set of parameters for direct comparison. 
Numerical results and their physical interpretations are discussed systematically in the following subsections.

\begin{figure}[b]
\centering
\includegraphics[width=1.0\columnwidth]{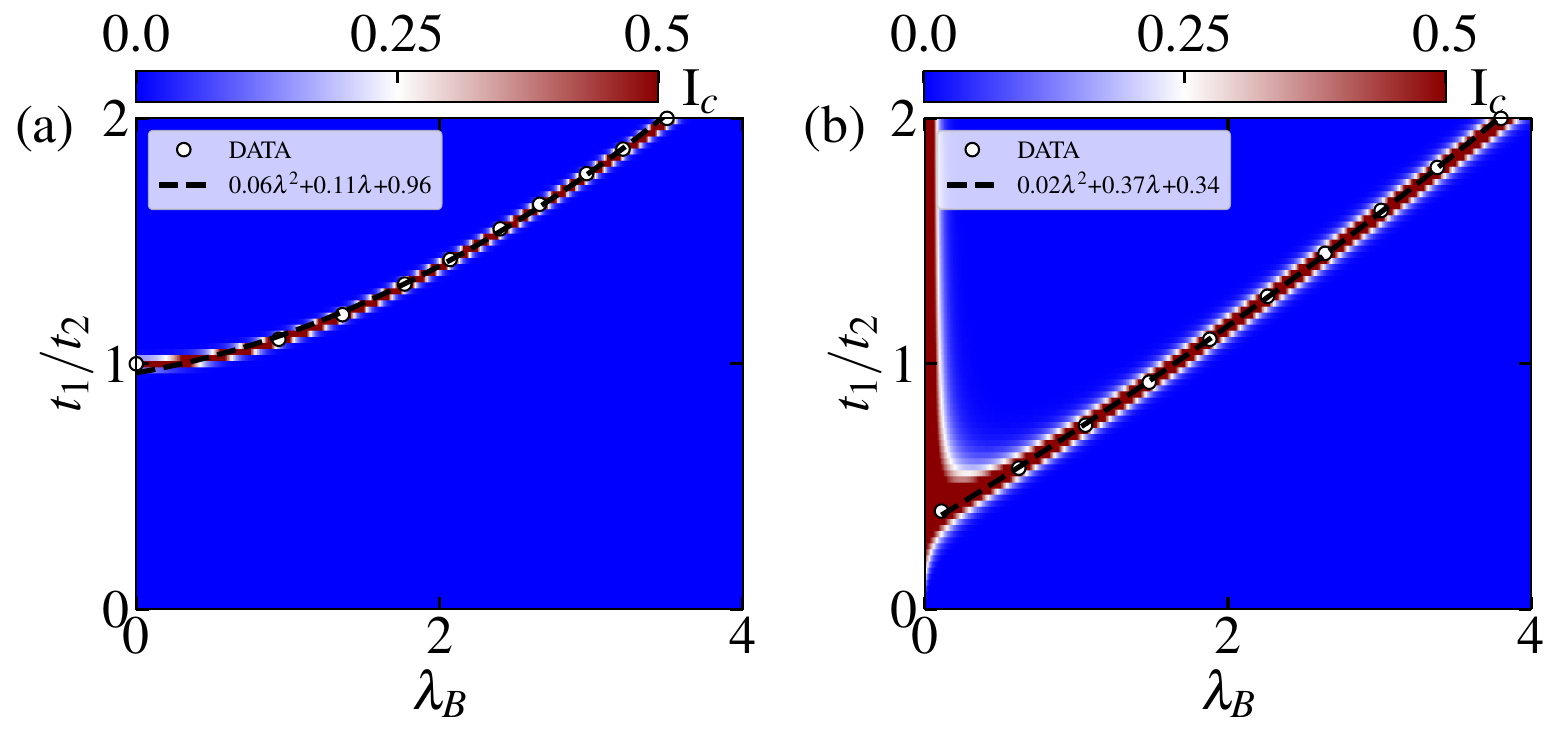}
\caption{Charge current $I_c$ plotted as functions of hopping dimerization $t_{1}/t_{2}$ and onsite potential modulation strength $\lambda_B$, for both the (a) half-filling and (b) special filling configurations. The color bar indicates the magnitude of $I_c$ which has been set between $0.0$ and $0.5$ for clarity. The parameter $\lambda_A$ is fixed at zero. 
The dashed curve denotes the fitted quadratic function with the current maximas (white circles).}
\label{fig:fig3}
\end{figure}

\begin{figure*}[t]
\centering
\includegraphics[width=2.0\columnwidth]{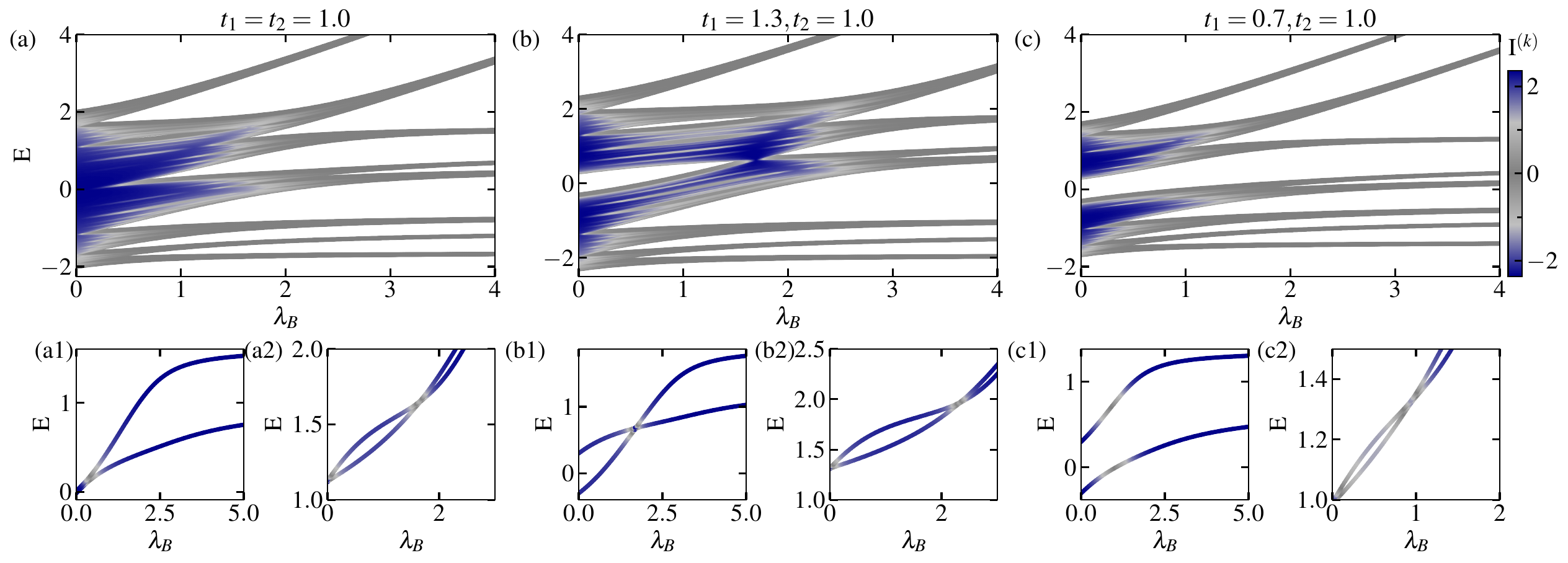}
\caption{Energy spectrum as a function of the onsite potential strength $\lambda_B$ with corresponding state current ($I^{(k)}$) for (a) $t_1=1.0$, (b) $1.3$ and (c) $0.7$. Here, the other parameters are $t_2=1.0$, $\lambda_A=0$, and system size $2F_n=288$. The subplots in the lower panel with indices 1 and 2 depict the energy spectrum near the half and special fillings, respectively.}
\label{fig:fig4}
\end{figure*}

\subsection{Non-interacting case ($U=0$)}
In this subsection we will first explore the behavior of the charge current complemented with the Drude weight and also studied the spin current for different fillings for the non-interacting case.
\subsubsection{Charge current and Drude weight}
First of all, we examine the behavior of the charge current ($I_c$) and the Drude weight ($D_w$) as functions of onsite modulation $\lambda_B$ in Fig.~\ref{fig:fig2} plotted as red squares and blue circles, respectively, for different dimerization strengths (a, d) $t_2=1.0$, (b, e) $t_2=1.3$ and (c, f) $t_2=0.7$. We obtain that for half-filled case (a-c), when $t_1 = t_2$, an initially finite $I_c$ sharply vanishes with increase in $\lambda_B$. However, for $t_1> t_2$, we obtain a sharp enhancement in current at a critical $\lambda_B$ although it vanishes at other values of $\lambda_B$. For $t_1 < t_2$, the current completely vanishes for all values of $\lambda_B$ considered. Similar features are also seen from the Drude weight $D_w$ for all the three cases. 
Interestingly, however, a distinct re-entrant behavior is observed in both $I_c$  and $D_w$ as a function of $\lambda_B$ at a special filling which depends upon the Fibonacci number, i.e, $(2F_{n-1} +F_{n-4})=199$
for all the hopping dimerization as shown in Fig.~\ref{fig:fig2} (d-e) while one can also observe this similar phenomena by varying $\lambda_A$ with filling $F_{n-1}=89$. 
In this case, the current initially decreases with $\lambda$, then re-emerges beyond a threshold, forming pronounced peaks. This re-entrant nature, which is absent for half-filling case, underscores a fundamental difference in transport mechanism governed by electron filling.
We also examine the sensitivity of the critical $\lambda$ for the maximum $I_{c}$ with respect to  $t_1/t_2$  for half filling in Fig.~\ref{fig:fig3}(a) and for special filling $(2F_{n-1} + F_{n-4})$
in Fig.~\ref{fig:fig3}(b) depicting a quadratic dependence for both the cases.

In general, for disordered systems, including those governed by the AAH model or Fibonacci chains, increasing the strength of the onsite-site potential tends to localize the states and suppress electronic transport, manifesting as a reduction in both current and Drude weight. However, certain variants of the AAH model, particularly those with asymmetric hopping, have been shown to exhibit transport enhancement under specific conditions due to resonance effects or mobility edges. In contrast, the present model incorporates a deterministic aperiodic modulation in the form of a Fibonacci sequence, rather than a quasiperiodic disorder, leading to distinct transport features. For the half-filled case, when the intra-cell hopping dominates ($t_1 > t_2$), we observe a non-monotonic response, both $I_c$ and $D_w$ initially increase with increasing $\lambda_A$ or $\lambda_B$, indicating a regime where moderate modulation enhances delocalization or coherence in transport. Beyond a critical strength, further increase in site energy suppresses transport, consistent with the expected localization-driven behavior. In the specially filled case, a distinct reentrant behavior emerges in both the $I_c$ and $D_w$, these quantities initially decrease with increasing site energy, but beyond a certain threshold, they begin to rise again. This non-monotonic transport response is a direct consequence of the underlying Fibonacci ordering in the site potential and is consistently observed across different system sizes that adhere to the Fibonacci sequence. 

\begin{figure}[b]
\centering
\includegraphics[width=1.\columnwidth]{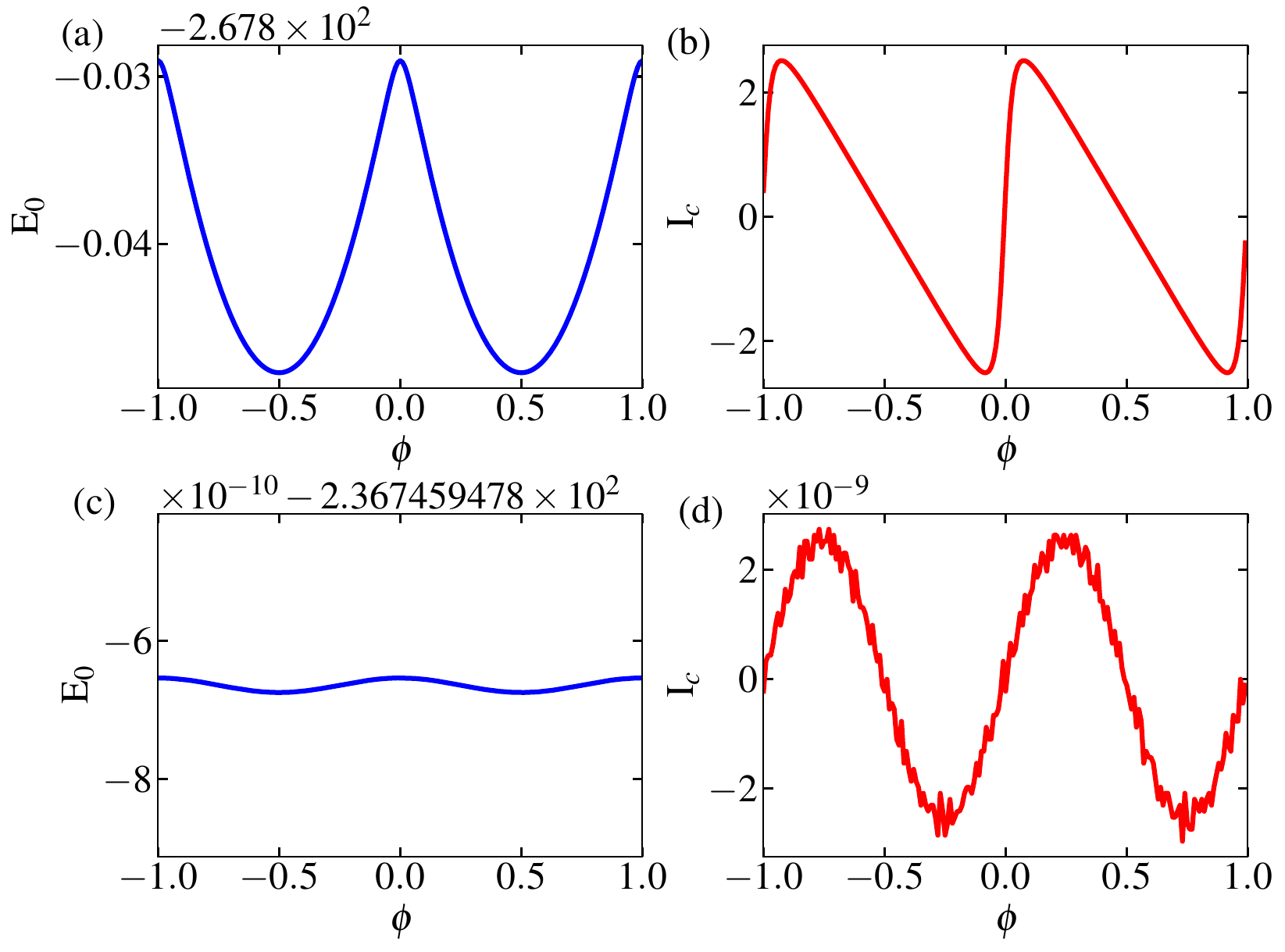}
\caption{Ground-state energy $E_g$ (a, c) and charge current $I_c$ (b, d) as a function of the magnetic flux $\phi$. The top panels correspond to $\lambda_B = 1.7$, while the bottom panels correspond to $\lambda_B = 2.0$. The other parameters are fixed at $t_1 = 1.3$, $t_2 = 1.0$, $\lambda_A = 0.0$, and system size $2F_n = 288$.
}
\label{fig:fig5}
\end{figure}

We also observe that the position of the peak in $I_c$ and $D_w$ is sensitive to the interplay between the hopping asymmetry and site energy modulation. Specifically, when varying $\lambda_A$, the relevant special filling corresponds to $F_{n-1}$, 
whereas for $\lambda_B$ variations, it is $(2F_{n-1} + F_{n-4})$, 
with $F_n$ denoting the $n^{\text{th}}$ Fibonacci number. These filling fractions stem from the recursive nature of the Fibonacci structure, which governs the spatial distribution of the potential landscape. Importantly, this reentrant behavior arises without the presence of quasiperiodic disorder, emphasizing that deterministic aperiodicity alone is sufficient to induce non-trivial transport phenomena. The results underscore how the interference effects and spatial correlations inherent to Fibonacci-type ordering can enhance coherence and enable tuning of transport properties through selective filling and site modulation.


To elucidate the influence of the onsite Fibonacci potential on current transport, we present in Fig.~\ref{fig:fig4} the energy spectrum as a function of onsite potential strength, with the color scale encoding the state current $(I^{(k)})$ associated with each eigenstate.
At weak disorder, the current is generally suppressed near the spectral edges and becomes more pronounced near the center of the band. For $\lambda_A = 0$ and $t_1 = t_2$, the system reduces to a standard tight-binding model with a continuous, gapless spectrum. However, when $t_1 \neq t_2$, a spectral gap appears even at zero potential strength, signifying an insulating phases. These correspond to the well-known topologically trivial and non-trivial insulators, depending on the dimerization pattern, highlighting the role of hopping asymmetry in determining the bulk electronic structure. The lower panels in each subfigure detail the evolution of eigenstates near half-filling and special filling. For $t_1 = t_2$, the energy levels around half-filling initially approach each other as $\lambda_B$ increases, indicating enhanced transport due to reduced excitation gaps. With further increase in potential strength, the spectra diverge, signaling a transition to an insulating state. At special filling, a reentrant behavior is observed, spectral convergence leads to enhanced current, followed by suppression and then a secondary enhancement as bands realign. This explains the non-monotonic trends seen in both $I_c$ and $D_w$ (compare Fig.~\ref{fig:fig2}(d)). A similar pattern arises for $t_1 > t_2$ and $t_1 < t_2$, where the convergence of spectra correlates with maximal current. Conversely, spectral divergence corresponds to diminished current, approaching insulating behavior. 

To gain deeper insight, we study how the ground-state energy and charge current vary with the applied magnetic flux $\phi$ at half filling. Fig.~\ref{fig:fig5} shows the ground-state energy $E_g$ and the corresponding current $I(\phi)$ for two values of the potential strength, i.e.,  $\lambda_B = 1.7$ and $2.2$. These values are chosen because they coincide with a peak and a dip in the Drude weight, as seen in Fig.~\ref{fig:fig2}(b). We obtain that for $\lambda_B = 1.7$, $E_g$ displays clear periodic oscillations with $\phi$ shown in Fig.~\ref{fig:fig5}(a), reflecting a strong flux sensitivity and finite current which looks like a saw-tooth behavior shown in Fig.\ref{fig:fig5}(b). However, when $\lambda_B$ is increased to $2.2$, these oscillations are strongly suppressed shown in Fig.~\ref{fig:fig5}(c), and $E_g$ becomes nearly independent of $\phi$ and thus the current is zero shown in Fig.~\ref{fig:fig5}(d).

 \begin{figure}[t]
\centering
\includegraphics[width=1.0\columnwidth]{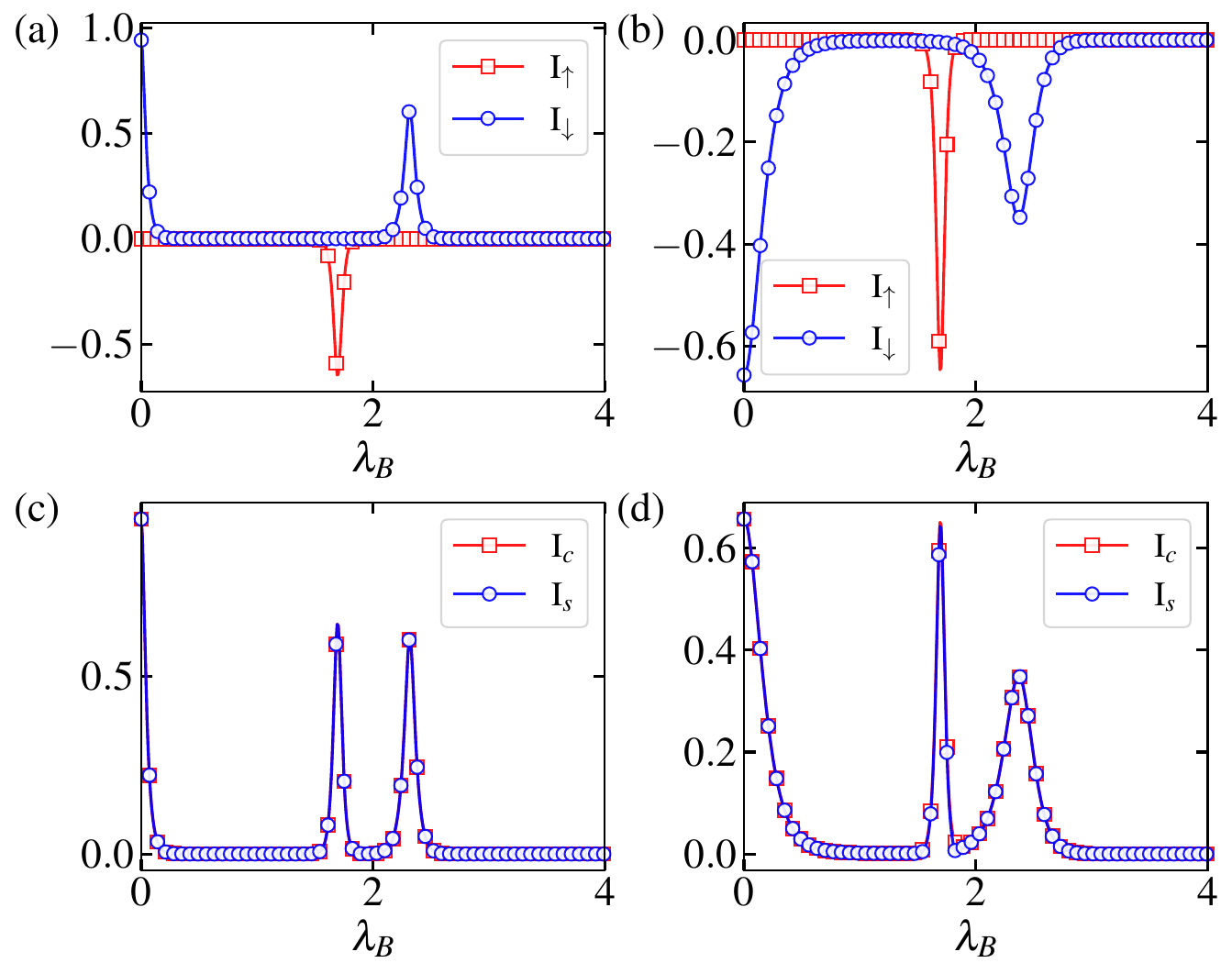}
\caption{(a) and (b) show the variation of the up- and down-spin currents, while panels (c) and (d) present the corresponding charge and spin currents as function of the site-energy strength.
In panels (a) and (c), $n_{\downarrow}=199$ $(2F_{n-1}+F_{n-4})$, while in panels (b) and (d), $n_{\downarrow}=110$ $ (2F_{n-2})$. In all cases, $n_{\uparrow}$ is fixed at $144$ with $t_{1}=1.3 > t_{2}=1$ and system size $2F_n=288$.}
\label{fig:fig6}
\end{figure}

\begin{figure}[b]
\centering
\includegraphics[width=1.0\columnwidth]{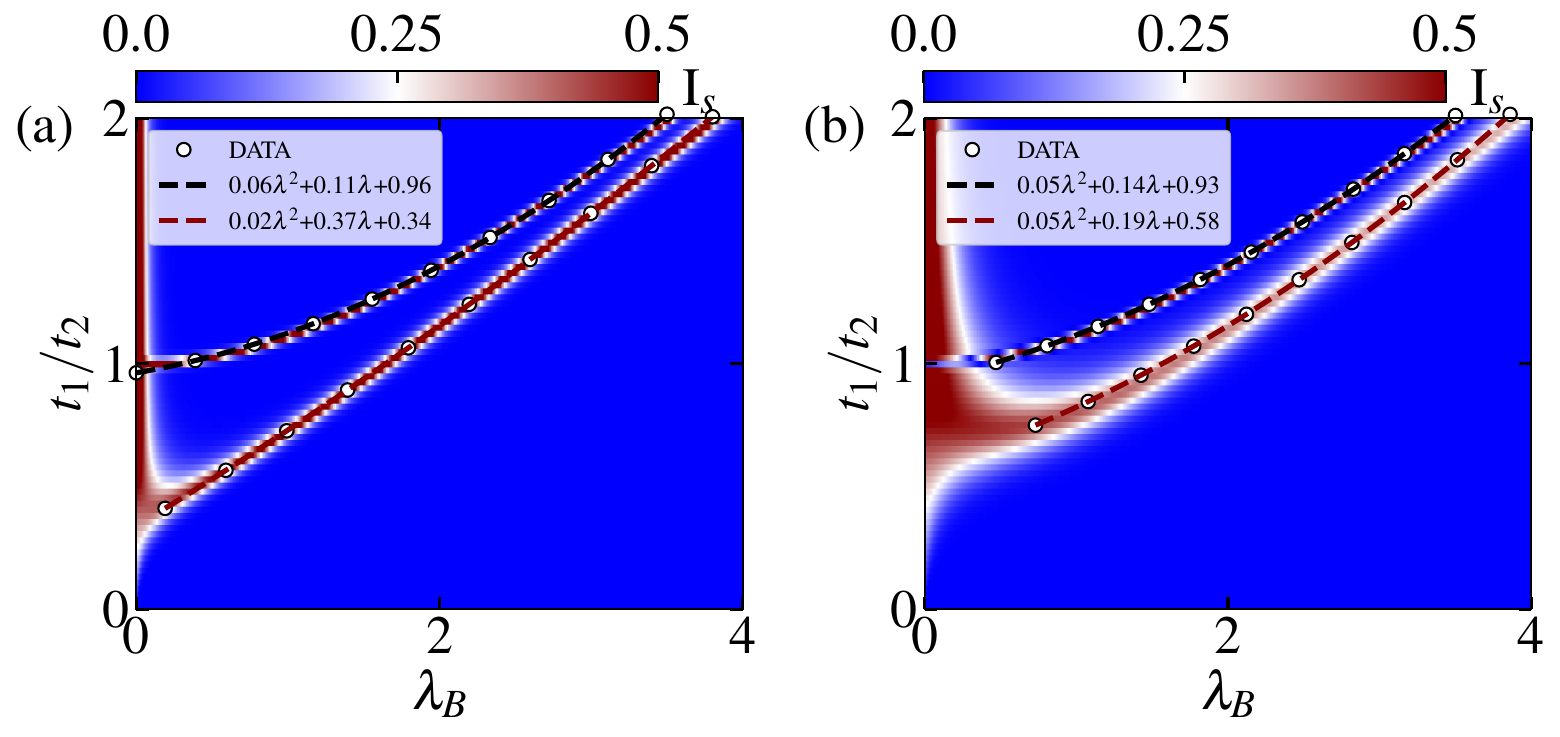}
\caption{Spin current $I_s$ plotted as functions of hopping dimerization $t_{1}/t_{2}$ and onsite potential modulation strength $\lambda_B$. Results are shown for 144 up-spin, and (a) 199 ($2F_{n-1}+F_{n-4})$ and (b) 110 $(2F_{n-2})$ down-spin. The color bar indicates the magnitude of $I_s$ which has been set between $0.0$ and $0.5$ for clarity. We set $\lambda_A=0$. The dashed curve denotes the fitted quadratic function with the current maximas.}
\label{fig:fig7}
\end{figure}

\begin{figure*}[t]
\centering
\includegraphics[width=2.0\columnwidth]{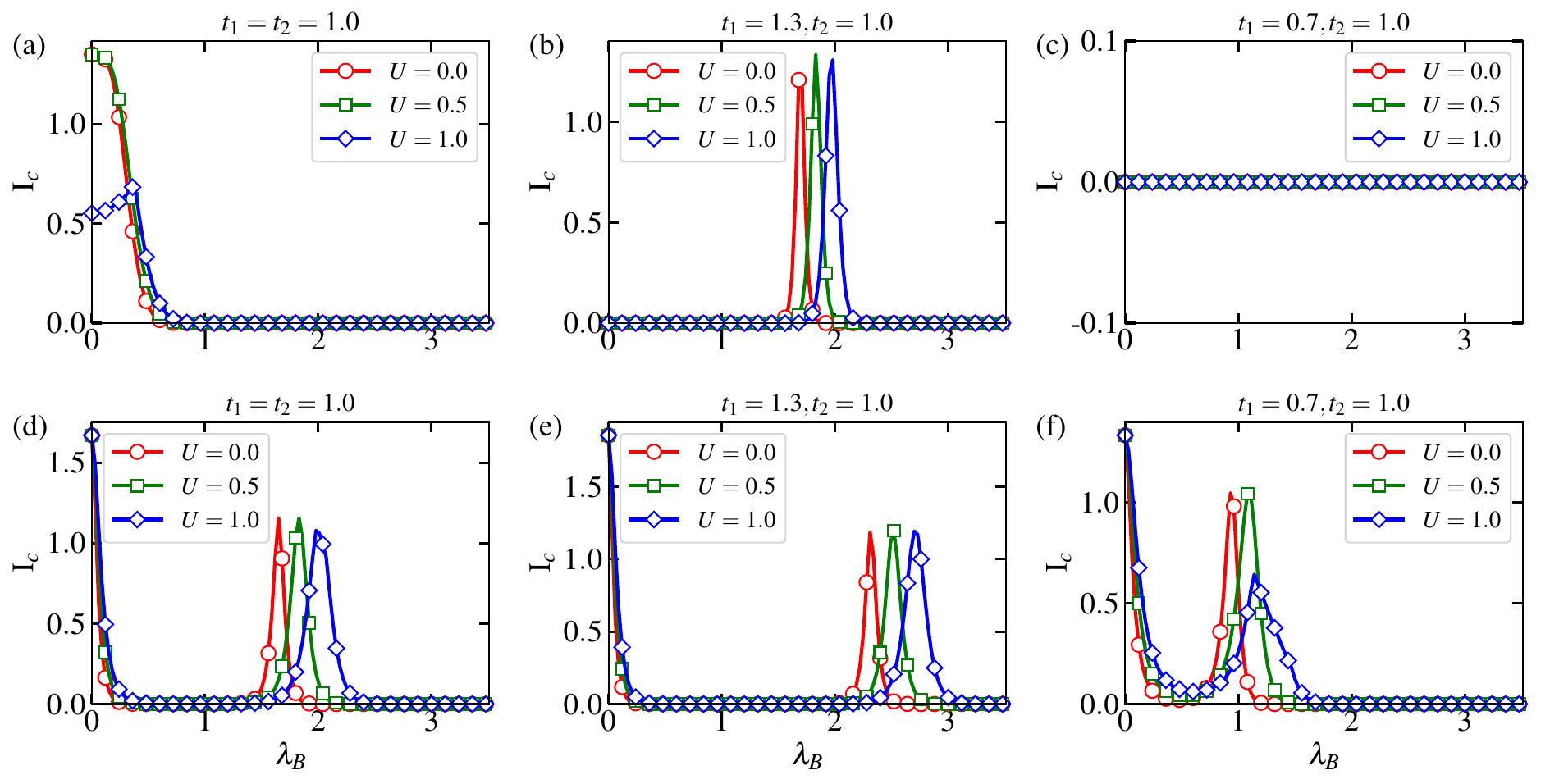}
\caption{
The charge current $I_c$ as a function of $\lambda_B$ for different values of interaction $U=0.0$ (red circles), $U=0.5$ (green squares) and $U=1.0$ (blue diamonds).
Upper panel (a-c) correspond to the half-filling case $(F_n=144)$  
whereas the lower panel (d-f) depict results for special filling case $(2F_{n-1} + F_{n-4}=199)$ for a system of size $2F_n=288$. 
The hopping amplitudes are set as $t_1 = 1.0$, $1.3$, and $0.7$, for left , middle, and right columns, respectively, with $t_2$ fixed at 1.0 with $\lambda_A=0$.
}
\label{fig:fig8}
\end{figure*}

\subsubsection{Spin current}

So far, we have analyzed the charge current properties for equal spin populations, as identical spin populations prevail, the spin current becomes zero.  We now turn to the spin imbalanced case to explore its impact on both spin and charge currents. Fig.~\ref{fig:fig6} illustrates these results by showing how the currents evolve with the site-energy strength $\lambda_B$ for two different down-spin fillings, odd ($n_\downarrow = 199$) and even ($n_\downarrow = 110$), while keeping the up-spin number fixed at $n_\uparrow = 144$. Panels (a) and (b) display the individual up- and down-spin currents, respectively, while panels (c) and (d) show the corresponding charge and spin currents. Note that in the equal-spin case as discussed previously, we observe a re-entrant pattern in the charge current as $\lambda_B$ varied. However, under spin imbalance,  this re-entrant behavior becomes richer. In this case, both charge and spin currents exhibit multiple re-entrant transitions as shown in Fig.~\ref{fig:fig6}(c) and (d). These arise directly from the imbalance between up- and down-spin populations, which modifies how the system responds to changes in the potential strength. Interestingly, the structure of these transitions depends  on whether the down-spin filling is odd or even, with the odd case displaying a more stable two-peak spin-current profile as shown in Fig.~\ref{fig:fig6}(c). To understand this origin of the multiple re-entrant phenomena, it is helpful to examine the band structure as shown in Fig.~\ref{fig:fig4}(b). The subfigures present the cases of half-filling [Fig.~\ref{fig:fig4}(b1)] and special filling [Fig.~\ref{fig:fig4}(b2)]. For the up-spin sector, the filling fraction matches the half-filling band diagram, whereas for the down-spin sector, the filling fraction is special. In this situation, a peak appears when the bands close to this filling fraction approach each other. Consequently, the corresponding spin-up and spin-down currents are shown in Fig.~\ref{fig:fig6}(a), from which both the spin and charge current behaviors can be seen in Fig.~\ref{fig:fig6}(c). 

To gain a deeper insight into this we present the spin current in the $t_1/t_2-\lambda_B$ plane for a fixed up-spin number $n_\uparrow = 144$, with odd ($n_\downarrow = 199$) and even ($n_\downarrow = 110$) down-spin fillings in Fig.~\ref{fig:fig7}(a) and (b), respectively. The color scale indicates the value of current.  
The diagrams show that spin current strongly depends on both the hopping dimerization and the potential strength. The multiple re-entrant transitions, where the current turns on and off repeatedly, are clearly visible. These features are more prominent for $t_1 > 1$, especially in both the even and odd filling case. Interestingly, for odd filling, additional re-entrant behavior appears below $t_1 < 1$, which is absent in the even case. Here, we also observe that the peaks exhibit a clear quadratic dependence in both cases.

\subsection{Interacting case}
In this subsection we investigate the effect of electron–electron interactions $U$ on the charge and spin transport properties.

\subsubsection{Charge current}

\begin{figure}[t]
\centering
\includegraphics[width=1.0\columnwidth]{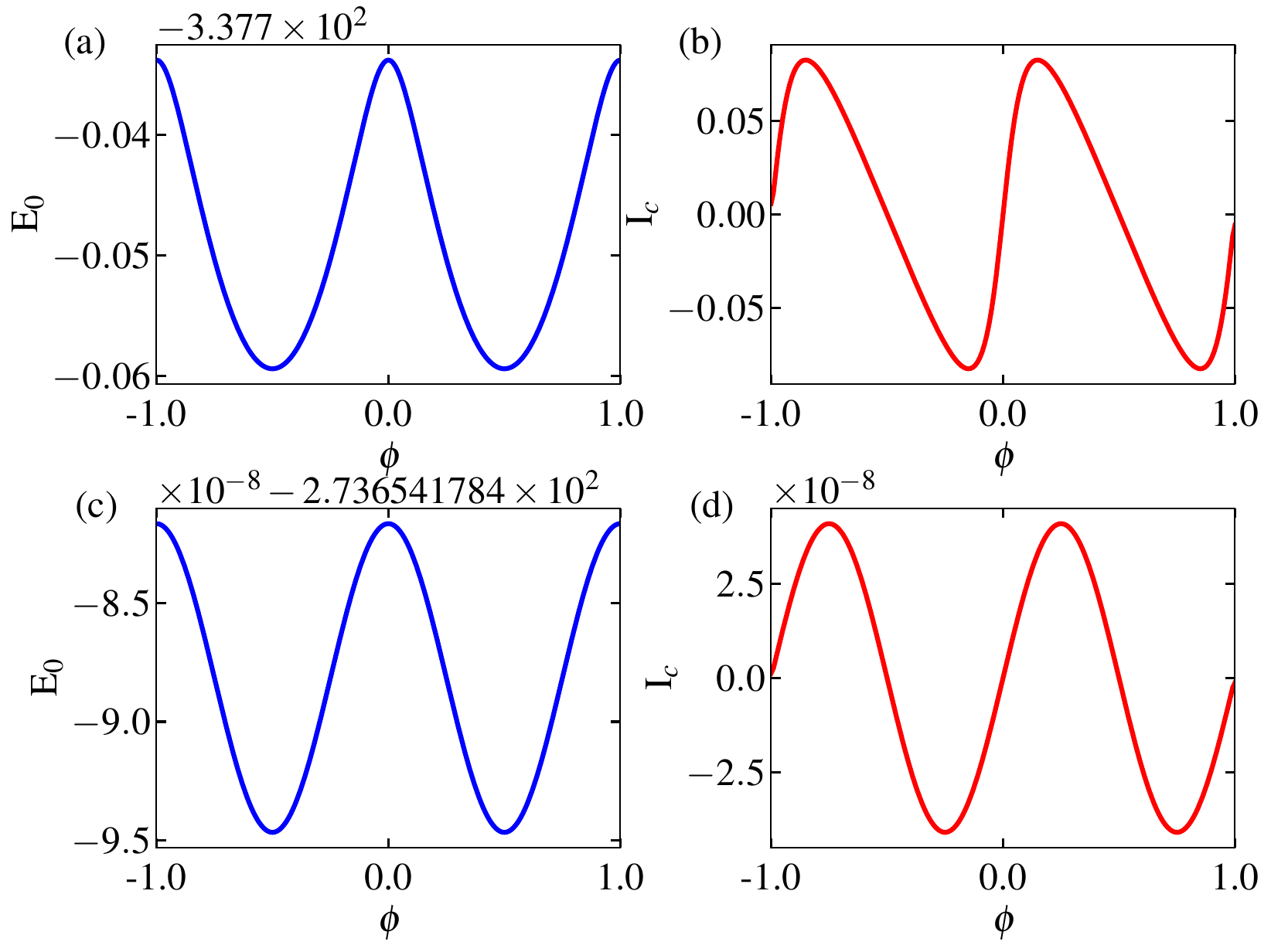}
\caption{Ground-state energy (a, c) and charge current (b, d) as functions of the magnetic flux $\phi$. The top panels correspond to $\lambda_B = 2.0$, while the bottom panels correspond to $\lambda_B = 2.5$. The other parameters are fixed at $t_1 = 1.3$, $t_2 = 1.0$, $\lambda_A = 0.0, U=1$, and system size $2F_n = 288$.
}
\label{fig:fig9}
\end{figure}

\begin{figure}[b]
\centering
\includegraphics[width=1.0\columnwidth]{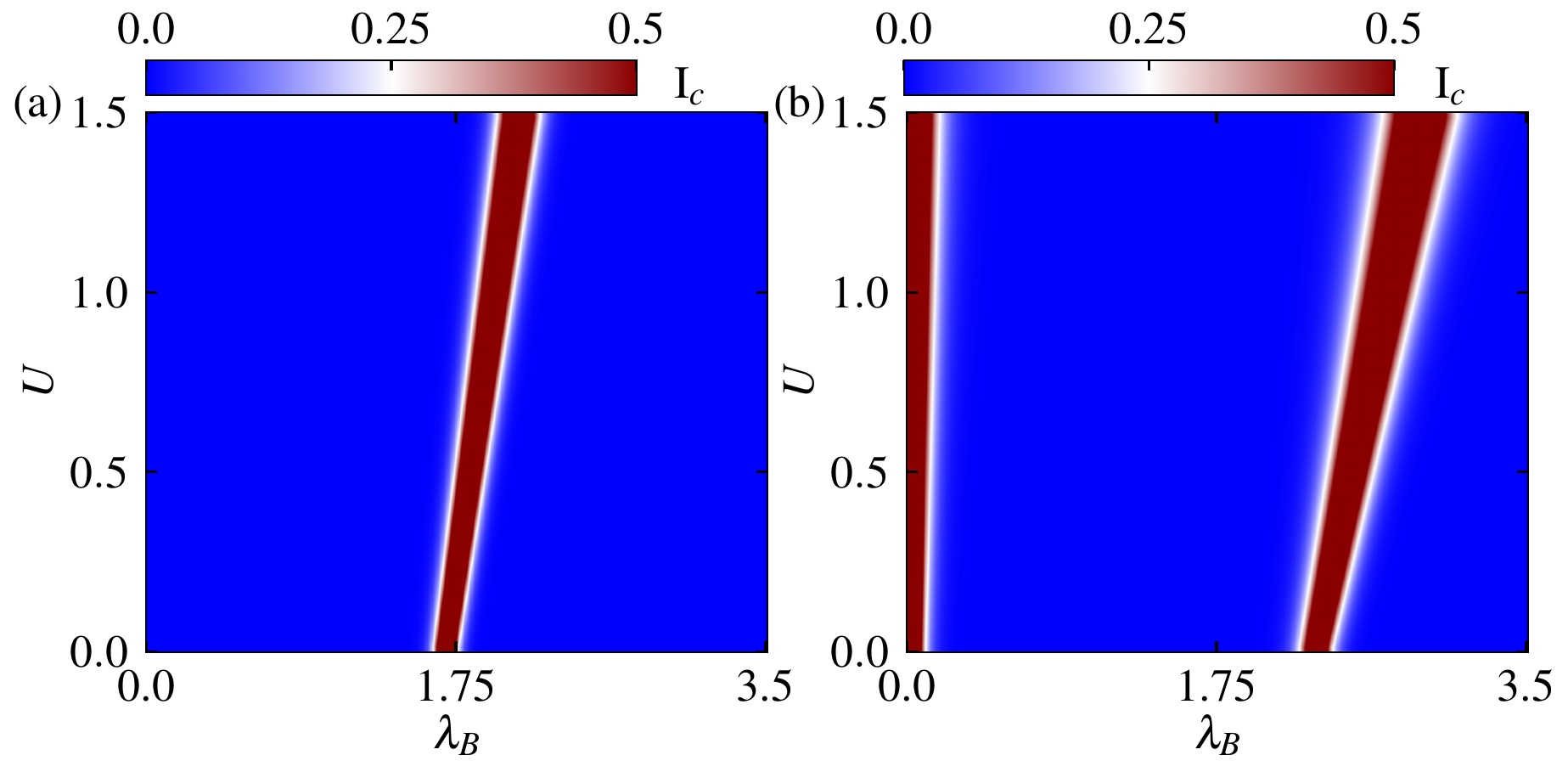}
\caption{Charge current $(I_c)$ as function of $\lambda_B$ and $U$ for (a) half-filling and (b) special-filling cases with $t_1=1.3 > t_2=1.0$ and $\lambda_A=0$. The color bar indicates the magnitude of $I_c$ which has been set between $0.0$ and $0.5$ for clarity.}
\label{fig:fig10}
\end{figure}

To study how transport is influenced by the interplay between the Fibonacci potential and electron interactions, we begin by analyzing the charge current as a function of the onsite potential strength $\lambda_{B}$. Fig.~\ref{fig:fig8} presents the results for three representative interaction strengths, $U=0.0$ (red circles), $U=0.5$ (green squares), and $U=1.0$ (blue diamonds). The figure compares both the half-filling case (top row) and special Fibonacci filling (bottom row) cases. At half-filling with equal hoppings ($t_{1} = t_{2}$), the current $I_c$ starts with a finite value but rapidly decreases as $\lambda_{B}$ increases (Fig.~\ref{fig:fig8}(a)). 
When the hopping is dimerized ($t_{1} > t_{2}$), $I_c$  shows sharp peaks at a critical value of $\lambda_{B}$ which shifts to larger values as $U$ increases (Fig.~\ref{fig:fig8}(b)).
On the contrary, for the dimerization of type $t_1<t_2$, $I_c$  is essentially zero for all the values of $U$ as shown in Fig.~\ref{fig:fig8}(c). For the special filling (bottom row of Fig.~\ref{fig:fig8}) the scenario completely changes. In this case, $I_{c}$ first falls with $\lambda_B$ but then reappears at larger $\lambda_B$, producing clear re-entrant peaks for both dimerized and uniform hopping cases. These peaks also shift slightly with $U$, but the overall pattern is less sensitive to hopping dimerization as is the case for half-filled system. At half filling, the system is close to a band-insulating or particle-hole symmetric point, so weak to moderate interactions slightly perturb the electron distribution, resulting in only a slow shift of the current peak with \(U\). In contrast, at the special filling, electrons occupy higher-energy states that are more sensitive to both Fibonacci modulation and interactions. Consequently, Hubbard correlations strongly redistribute the electrons, leading to a more rapid shift of the current peak with increasing \(U\).

\begin{figure}[t]
\centering
\includegraphics[width=1.0\columnwidth]{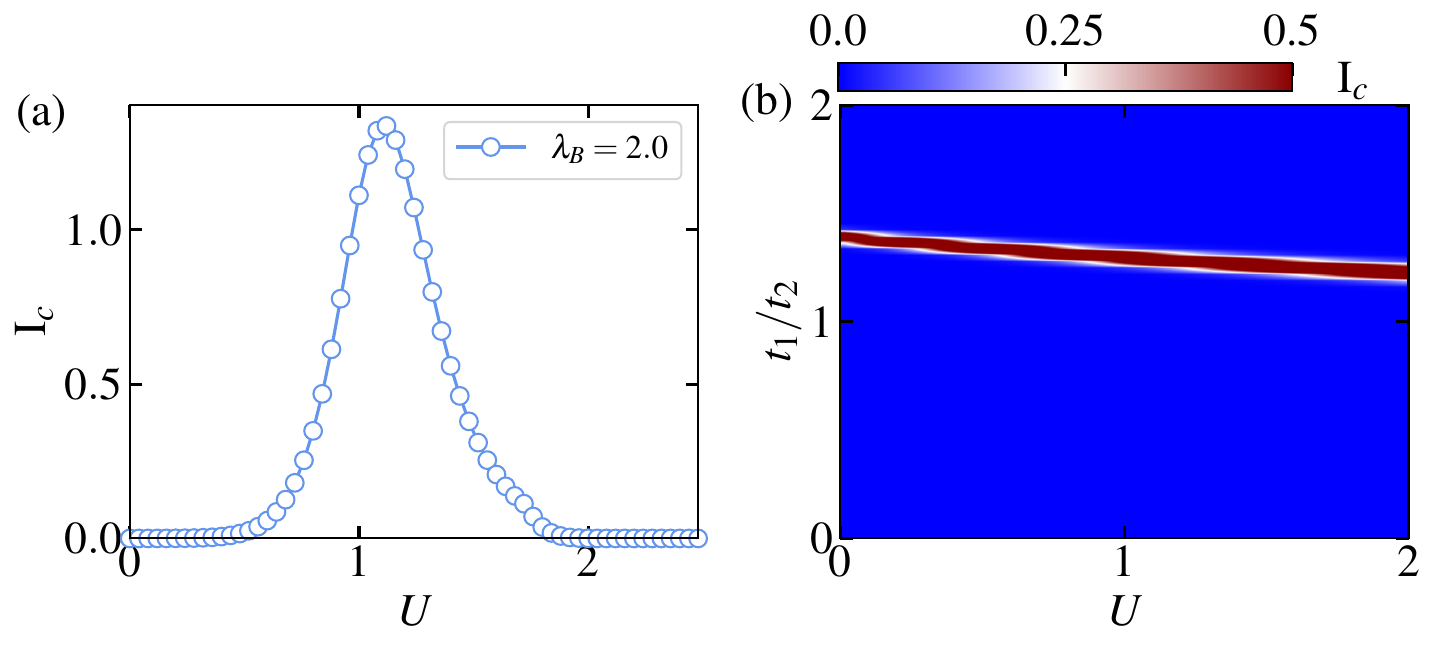}
\caption{(a) Charge current ($I_c$)as a function of interaction strength at half filling case. (b) Variation of $I_c$ with hopping dimerization $t_1/t_2$ and interaction strength $U$ for $\lambda_B=2$. The color bar indicates the magnitude of $I_c$ which has been set between $0.0$ and $0.5$ for clarity. We assume $\lambda_A=0$.  }
\label{fig:fig11}
\end{figure}

To further understand this behaviour, we examine the dependence of the ground-state energy $E_{g}$ and the corresponding charge current $I_c$ on the applied magnetic flux $\phi$ at half filling. Fig.~\ref{fig:fig9} displays the results for two representative values of the site-energy modulation, $\lambda_{B} = 2.0$ and $2.5$. For $\lambda_{B} = 2.0$, $E_g$ exhibits pronounced periodic oscillations with $\phi$, as shown in Fig.~\ref{fig:fig9}(a), signaling strong flux sensitivity characteristic of finite current in the system. The associated current profile in Fig.~\ref{fig:fig9}(b) reflects this behavior, with large amplitude oscillations. In contrast, when $\lambda_{B}$ is increased to $2.5$, the oscillations in $E_{g}$ are markedly suppressed [Fig.~\ref{fig:fig9}(c)], rendering it nearly flux-independent. Correspondingly, the charge current in Fig.~\ref{fig:fig9}(d) is strongly diminished and thus the system behaves as an insulator.
 
We now depict the variation of $I_c$ as a function of $U$ and $\lambda_B$ in Fig.~\ref{fig:fig10} for both half-filling in panel (a) and special-filling in panel (b), with hopping dimerization fixed as $t_{1} (1.3) > t_{2} (1.0)$. At half filling [Fig.~\ref{fig:fig10}(a)], the reddish current peak near $\lambda_{B} \approx 1.75$ for $U=0$ broadens as $U$ increases, showing that interactions enlarge the parameter window where transport is sustained. In the special-filled case [Fig.~\ref{fig:fig10}(b)], we observe re-entrant behavior in current, where both the primary and secondary peaks broaden, with the secondary peak is more sensitive to $U$ than that of the primary peak. This indicates that transport at special fillings is more strongly affected by the interaction strength. In the non-interacting case, the current is dictated solely by the interplay between hopping correlation and site-energy modulation, leading to rapid decay when localization dominates. Introducing Hubbard interaction $U$ modifies the electron distribution and partially counteracts Fibonaci sequence induced localization. As a result, the current decays more slowly, and the peak broadens, reflecting enhanced transport over a wider parameter range of $U$.

So far, we have seen that tuning the modulation strength $\lambda_B$ can enhance the current. To probe this effect further, we now fix $\lambda_B=2.0$, where the current is essentially suppressed (as shown in Fig.~\ref{fig:fig8}(b)), and investigate how the system responds when the interaction strength $U$ is increased. The corresponding results are presented in Fig.~\ref{fig:fig11}.
Panel (a) shows the variation of the current with $U$ at half filling. The behavior is clearly non-monotonic, in the non-interacting limit ($U=0$) the current is nearly suppressed, then it rises sharply with increasing $U$, reaching a pronounced maximum around $U \approx 1$, before decreasing again at larger $U$. This dome-shaped profile demonstrates that moderate interactions enhance transport by partially delocalizing states that are otherwise strongly localized by the Fibonacci potential. However, strong interactions drive the system back toward insulating behavior, causing the current to diminish.
Panel (b) further illustrates this effect by mapping the current in the parameter space spanned by the hopping ratio $t_1/t_2$ and the interaction strength $U$. The reddish region highlights the window where the current is maximized, confirming that the interaction-driven enhancement of transport is robust even under moderate hopping asymmetry. The slight downward tilt of this reddish region indicates that increasing hopping asymmetry shifts the optimal $U$ to smaller values.
Overall, these results clearly show that electron interactions can restore transport in regimes where the Fibonacci modulation alone would render the system insulating.

\begin{figure}[t]
\centering
\includegraphics[width=1.0\columnwidth]{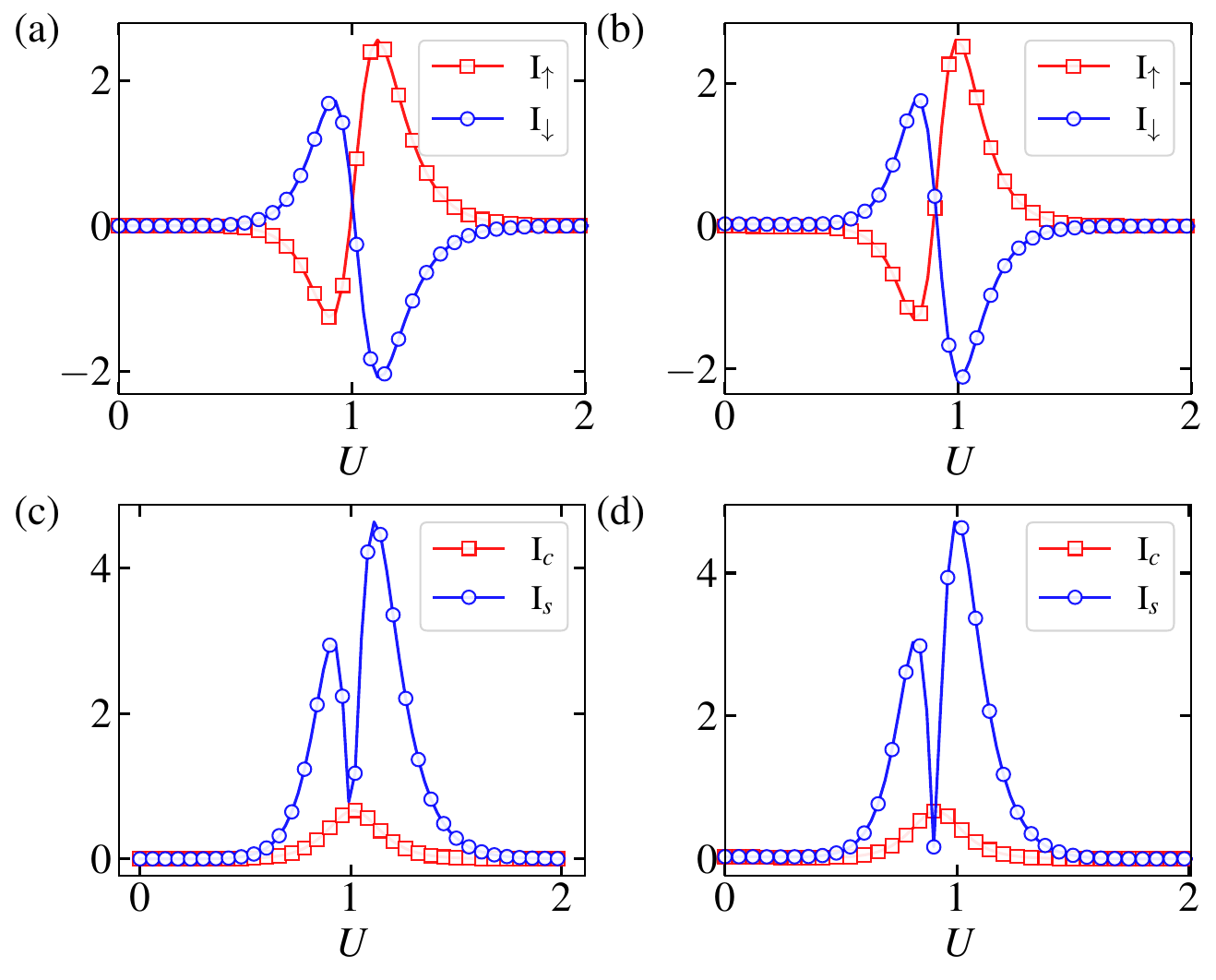}
\caption{Variation of $I_\uparrow$ and $I_\downarrow$ with $U$.  (a) and (b) correspond to $n_{\downarrow}=199$ and $n_{\downarrow}=110$, while (c) and (d) depict the charge and spin currents for the same cases. The parameters are fixed at $t_{1}=1.3$, $t_{2}=1.0$, $\lambda_{B}=2.0$, $n_{\uparrow}=144$, with a total system size $2F_n=288$.
}
\label{fig:fig12}
\end{figure}

\subsubsection{Spin current}

In this subsection, we shift our focus from the charge current, studied so far, to the role of spin imbalance in generating spin and charge current in the presence of interaction. Here, we fix the number of up-spin electrons at $n_\uparrow = 144$ and vary the number of down-spin electrons between $n_\downarrow = 199$ and $n_\downarrow = 110$. The corresponding variations of the spin-resolved currents with interaction strength $U$ are shown in Fig.~\ref{fig:fig12}(a) and (b). A distinct feature appears around $U \approx 1$, where the up-spin and down-spin currents exhibit opposite signs, whenever the up-spin current becomes positive, the down-spin current turns negative, and vice versa. This reversal indicates the onset of a spin-flip like behavior, arising from the combined effect of spin imbalance and interaction.  
To further clarify this, we present in Figs.~\ref{fig:fig12}(c) and (d) the charge current ($I_{\text{c}}$) and spin current ($I_{\text{s}}$) for the same cases. The spin current shows pronounced oscillatory variations with $U$, displaying multiple peaks and sign changes, which highlight its sensitivity to both interaction and spin imbalance. In contrast, the charge current follows a much simpler trend, with a single broad peak around $U \approx 1$. Notably, there are regions of $U$ where the charge current nearly vanishes while the spin current remains finite. This demonstrates that spin imbalance in the presence of interactions can drive a finite spin current even when the net charge current is suppressed.

\begin{figure}[t]
\centering
\includegraphics[width=1.0\columnwidth]{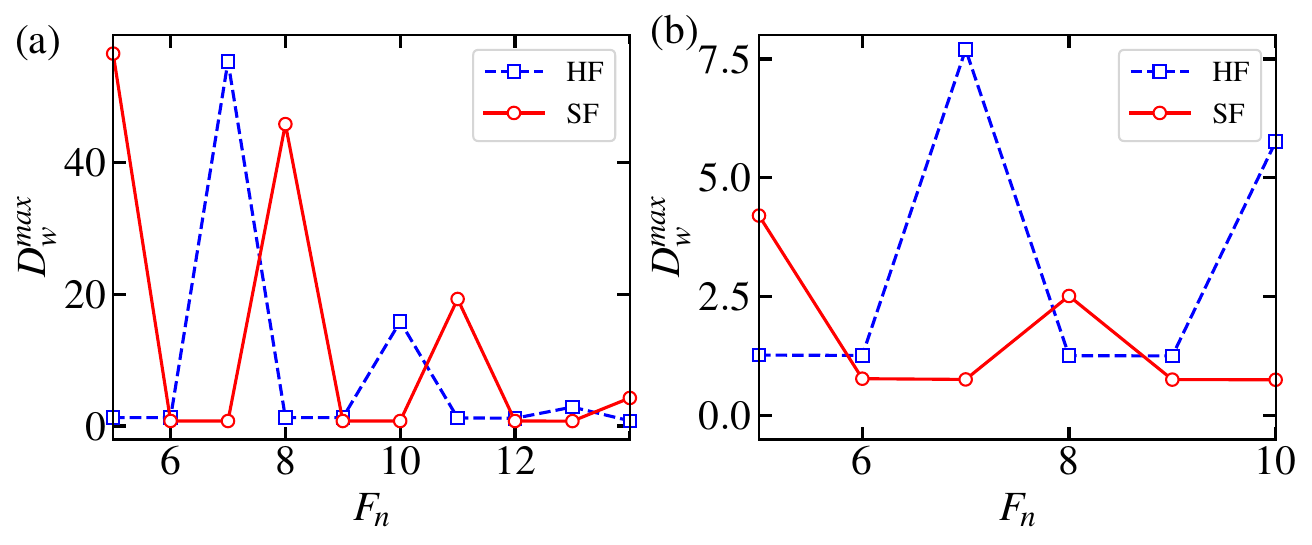}
\caption{The Drude weight corresponding to the Fibonacci sequence is shown for the half-filled and specially filled cases, respectively. The hopping ratio is fixed at $t_1 > t_2$, and $\lambda_A$ (or $\lambda_B$) is set to the value at which the peak response occurs. We perform a scan over $\lambda_B$ and identify the maximum Drude weight within the regime that exhibits a single peak for HF and a re-entrant peak for SF. (a) and (b) are for $U=0$, and $U=1$ respectively. Owing to numerical convergence limits in the mean-field approach, we consider up to $F_{n}=10$ for $U=1$. }
\label{fig:fig13}
\end{figure}

\subsubsection{System size dependence}

We now examine how the Drude weight $D_w$ changes as the system size increases for the both non-interacting ($U=0$) and interacting cases ($U=1$), with the sizes chosen according to the Fibonacci sequence
We compare two filling conditions, half-filling (HF, red circles with solid line) and special filling (SF, blue squares with dashed line). We obtain that the maximum Drude weight ($D_w^{max}$) does not simply vanish with larger system sizes but instead follows an alternating pattern as shown in both the Fig.~\ref{fig:fig13}(a) and (b). At certain Fibonacci sizes, the Drude weight is high for HF while remains low for SF, and at the next to next size the situation is reversed. This flip–flop behavior produces a zigzag trend, where the peak of one curve aligns with the dip of the other. In addition, there is a particular system in between this where the two curves nearly have low values. In the presence of interactions, while the overall magnitude is suppressed, the characteristic flip–flop pattern remains robust. Overall, the figure highlights that the $D_w$ depends on both the system size and the filling condition.

\section{Conclusions}
This study explores how a Fibonacci sequence of site potentials, together with dimerized hopping amplitudes, influences the transport and magnetic response of a one-dimensional electronic system in ring geometry. We focus on charge current, spin current, and Drude weight, considering both non-interacting and interacting cases. For non-interacting case at half-filling, we observe the appearance of distinct peaks in the charge current and Drude weight. Interestingly, when the system is tuned to a special filling, the current shows a re-emergence with varying potential strength. Furthermore, in the presence of spin imbalance, multiple re-entrant peaks are found, highlighting rich transport behavior. To capture the overall dependence on hopping dimerization and potential strength, we present a complete picture. We also investigate the role of interactions within a mean-field framework. Our results show that the charge current remains robust even in the interacting case, with the re-entrant peak structure largely preserved. A key finding is that, under specific potential strengths, the current first increases and then decreases as the interaction strength is varied. In the presence of spin imbalance and interactions, we find a finite spin current can persist even when the net charge current is suppressed. 

Our study reveals unusual behaviour of charge and spin currents with Fibonacci type potential strengths, hopping dimerization, and inter-particle interaction. Our findings suggest possible ways to tune charge and spin currents in mesoscopic systems, which could be useful for designing interaction-driven, low-dissipation spintronic devices. In the future, one can also study finite-size effects, disorder, temperature dependence, and time-dependent driving to gain deeper insights into transport and control of such systems.

\section{Acknowledgment}
T.M. acknowledges support from Science and Engineering Research Board (SERB), Govt. of India, through project No. MTR/2022/000382 and STR/2022/000023.

\bibliography{ref}

\end{document}